\documentclass[prl,floatfix,showpacs,amsmath,twocolumn]{revtex4}
\usepackage{amssymb}
\usepackage{graphicx}

\begin{document}

\newcommand{\be}{\begin{eqnarray}}
\newcommand{\ee}{\end{eqnarray}}
\newcommand{\bea}{\begin{eqnarray}}
\newcommand{\eea}{\end{eqnarray}}
\newcommand{\bma}{\begin{subequations}}
\newcommand{\ema}{\end{subequations}}
\newtheorem{lemma}{Protocol}
\def\ket #1{\vert #1\rangle}

\def\lR{l^2_{\mathbb{R}}}
\def\RR{\mathbb{R}}
\def\E{\mathbf e}
\def\D{\boldsymbol \delta}
\def\S{{\cal S}}
\def\T{{\cal T}}
\def\dd{\delta}
\def\one{{\bf 1}}

\title{Ensemble Quantum Computation with atoms in periodic potentials}

\author{K. G. H. Vollbrecht$^1$, E. Solano$^{1,2}$, and J. I. Cirac$^1$}
\affiliation{$1$ Max-Planck Institut f\"ur Quantenoptik,
Hans-Kopfermann-Str. 1, Garching, D-85748, Germany \\
$^2$Secci\'{o}n F\'{\i}sica, Departamento de Ciencias, Pontificia
Universidad Cat\'{o}lica del Per\'{u}, Apartado 1761, Lima, Peru}

\date{\today}

\begin{abstract}
We show how to perform universal quantum computation with atoms
confined in optical lattices which works both in the presence of
defects and without individual addressing. The method is based on
using the defects in the lattice, wherever they are, both to
``mark'' different copies on which ensemble quantum computation is
carried out and to define pointer atoms which perform the quantum
gates. We also show how to overcome the problem of scalability on
this system.
\end{abstract}

\maketitle

Neutral atoms confined in (quasi) periodic optical potentials and
manipulated by lasers provide us with one of the most promising
avenues to implement a quantum computer or to perform quantum
simulations \cite{Sciencereview}. For example, a Bose--Einstein
Condensate can be loaded in an optical lattice achieving almost
unit occupation per lattice site through a superfluid--Mott
insulator quantum phase transition \cite{bosonsol4}. A universal
set of quantum gates can then be implemented by individual laser
manipulation and inducing cold collisions between the atoms
\cite{Jaksch7}. In several remarkable experiments, all these
phenomena have been observed \cite{Greiner5,Mandel10}.

At the moment, quantum computation with atoms in optical lattices
is hindered by three major obstacles: (1) Lack of addressability;
(2) Presence of defects; (3) Uncontrolled number of atoms. The
first obstacle is due to the fact that the separation between
atoms is of the order of an optical wavelength (that of the laser
which creates the confining potential) so that in order to address
them with a laser one has to focus it close to (or even beyond)
the diffraction limit. A possible way to circumvent this problem
consists of using optical superlattices \cite{super}, or other
optical micro traps \cite{micro}, in which the separation between
atoms increases. Quantum gates in these set--ups, however, may
become harder than in the standard optical lattice. The second
obstacle occurs due to the fact that there always exist sites that
have either no atom or more than one. A single defect will
unavoidably spoil any quantum computation, and may also have
important consequences in quantum simulations. In present
experiments one can estimate that the number of defects is
relatively high \cite{vazi}. This last obstacle can be to a very
large extent overcome by a filtering process \cite{CirZo}, where
the lattice sites in which there is more than one atom are emptied
until a single atom remains there. Alternatively, one can define
collective qubits independent of the number of atoms per sites
\cite{GaCi}. Both procedures should avoid situations in which
there exist a defect with no atom present. Finally, the number of
atoms which form the quantum computer must be well defined since,
otherwise, when performing quantum gates the rest of the atoms
will act as an environment.

In this paper we introduce a novel method of performing quantum
computations in optical lattices (or, more generally, periodic
potentials) which circumvents the above mentioned obstacles. One
of the fundamental ideas of our method is to use defects (which
are delocalized in the lattice) in order to mark the atoms that
build the quantum computer and to break the translational symmetry
in order to obtain addressability. Note that we do not know where
the defects are, but their only presence (wherever they are) is
sufficient for our purposes. On the other hand, the defects allow
us to build ``pointer'' sites, also delocalized, which will be
used to perform a universal set of quantum gates. Note also that
since there will be several defects in the atomic sample, we will
have several quantum computers running in parallel, randomly
distributed all over the optical lattice. This situation resembles
the ensemble quantum computation set--up \cite{GeCh}, and in fact
some of the ideas developed there can be directly incorporated in
our method to make it more efficient. As we will show, the method
alone suffers, parallel to what happens in ensemble quantum
computation, from the scalability problem. Here we will also
present a method to overcome it and to make the present proposal
scalable. Note that even though our method is developed for atoms
in optical lattices, some of its ideas may also apply to very
different implementations where similar obstacles are present.

We consider a three dimensional periodic potential in which atoms
are loaded. The atoms have three internal states,
$|a\rangle,|b\rangle$ and $|p\rangle$. The first two will later on
store the qubit, whereas the third one will be used by the
pointer. We will consider each 1D lattice independently, i.e. we
assume that tunnelling is switched off for all times along the $y$
and $z$ directions. Thus, we can effectively reduce the system to
a set of 1D periodic lattices. We will use a second quantization
description of the states; that is, for each lattice site $k$ we
will write a state $|m_k,m_k',n_k\rangle_k$, where $m_k,m_k'$, and
$n_k$ are natural numbers that indicate the occupation number of
levels $|a\rangle,|b\rangle$ and $|p\rangle$, respectively. Thus,
a typical state of one 1D lattice will be
 \be
 \label{state_general}
 \otimes_{k} |m_k,m_k',n_k\rangle_k.
 \ee

We will assume that four kind of basic operations are available.
These operations act in exactly the same way in each lattice site,
since we do not assume that the sites can be individually
addressed. On the other hand, they are based on physical processes
which have been demonstrated in the current experimental set--ups:

{\bf (i) Particle transfers} in between internal states. We will
consider two kinds: {\bf (i.1)} Those in which an integer number
of particles  is transferred between states $a$ and $p$. For
example,
 \be
 U_{m,n}^{m+x,n-x}: |m,0,n\rangle \leftrightarrow |m+x,0,n-x\rangle,
 \ee
where $x$ is an integer. Note that  for the unitary operator which
describes this process at each site holds
$U_{m,n}^{m+x,n-x}=U_{m+x,n-x}^{m,n}$. Another example that we
will use later on will be $W:|1,0,1\rangle \leftrightarrow
|0,1,1\rangle$. These operations can be carried out using the
blockade mechanisms which is present due to atom--atom
interactions \cite{CirZo}. {\bf (i.2)} Those which generate
superpositions. For example, $V=\exp(-i H_{ab} \pi/8)$, which acts
only on the $a$ and $b$ levels, with $H_{ab} = \hat{a}^\dagger
\hat{b}+\hat{b}^\dagger \hat{a}$, where $\hat a$ and $\hat b$ are
the annihilation operators for particles in states $\ket{a}$ and
$\ket{b}$, respectively. These operations can be easily carried
out using laser or rf fields.

{\bf (ii) Collisional shifts}: They are due to the interactions
between particles in the states $a$ and $p$. For example, the
unitary operation $C(\varphi):|1,0,1\rangle \leftrightarrow
e^{i\varphi}|1,0,1\rangle$ can be obtained by waiting the
appropriate time \cite{Jaksch7}.

{\bf (iii) Lattice shifts}: We denote by $S_x$ the operations
which shift the pointer states $x$ steps to the right. For
example, $S_{-1}$ transforms the state in
Eq.~(\ref{state_general}) to $\otimes_{k}
|m_k,m_k',n_{k+1}\rangle_k$. They can be realized by changing the
intensity and polarization of the lasers \cite{Jaksch7,Brennen8}.

{\bf (iv) Emptying sites}: All atoms in internal states $a$ or $p$
are thrown away. This can be done, for example, by switching off
the lattice potential for the corresponding internal state. We
will denote them by $E_b$ or $E_p$ and they transform the state in
Eq.~(\ref{state_general}) into $\otimes_{k} |m_k,m_k',0\rangle_k$,
and $\otimes_{k} |m_k,0,n_k\rangle_k$, respectively.

Initially, all atoms are in the internal state $|a\rangle$,
distributed along the lattice according to some probability
distribution, i.e. the state will be a mixture of states in the
form of Eq.~(\ref{state_general}) with $m_k'=n_k=0$. Thus, the
goal is to show how with these random states and with the
operations which are available in the lab, and that do not require
addressing, we can perform arbitrary quantum computations. This
will be achieved in two steps. First there will be a preparation
step, and then a computation step. At the end we will show how to
include an additional step to make the system scalable.

In the preparation stage of our protocol, only states $a$, and $p$
will be occupied. Thus, we will simplify our notation denoting
$|m,n\rangle:=|m,0,n\rangle$. Moreover, the states that we will
use now will be product states, i.e. of the form \be
\label{states} |m_1,n_1\rangle \otimes |m_2,n_2\rangle \ldots
\otimes |m_N,n_N\rangle \ee where we have not included the
subscript $k$ to simplify the notation. This step starts out by
reducing all occupation numbers $>2$ to two (Fig.\ 1). This is
done by applying the operation $U_{x,0}^{2,x-2}$ first and then
$E_p$, and then repeating those actions for $x>2$ (up to some
value of $x$ in which we are confident that no site with this
number of particles is present).

The next step is to ``format'' the lattice. We produce several
areas, randomly distributed across the lattice, with exactly $n$
neighboring sites having a single atom in $a$ and one site at the
right edge with two atoms, one in $p$ and the other in $a$ (see
Fig.\ 1). In order to accomplish this, we have to keep only the
areas in which initially there are $n$ neighboring two-atom sites
and a one-atom site at the edge. The rest of the atoms are thrown
away, and then we manipulate the remaining atoms to obtain the
desired states. The sites in which initially there was a single
atom that has survived will now contain the ``pointers'' (the
extra atom in level $p$). This atom will then be used to perform
the quantum gates.

\begin{figure}[t]
\centering
\includegraphics[width=85mm]{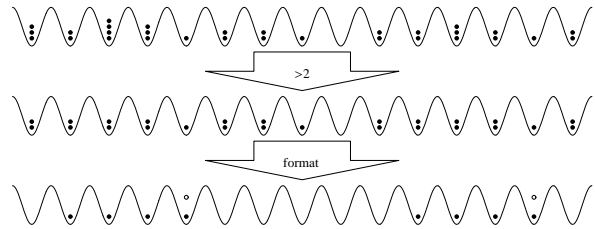}
\label{fig1} \caption{First, the sites with more than 2 atoms are
depopulated. The ``format'' step produces sites with 2 atoms in
levels $a$ and $p$, surrounded by a  ``reserved area'' to their
left which contains exactly $n$ sites with a single atom.}
\end{figure}

First, we change the internal states of the 1--atom sites from $a$
to $p$. These atoms are now called the ``pointers''. They will be
essential to create the quantum computers in the lattice. Each of
those atoms mark the position where we try to establish one of
those quantum computers. We want that such a pointer survives
during the following protocol if it has on its left at least $n$
sites with exactly 2-atoms in each. We thus proceed as follows. We
shift the pointer one lattice site to the left. We transfer the
pointer atom to the state $a$ iff there are two atoms in that site
by applying $U_{2,1}^{3,0}$. By emptying the internal states $p$
we delete all pointers which fail to have a 2-atom site next to
their starting position. Then we raise the pointer again by
$U_{2,1}^{3,0}$. By repeating this procedure for the next $n-1$
sites we delete all pointers that fail to have $n$ 2-atom sites on
the left of their starting position. Note that this also implies
that every pointer in one of the $n$ sites on the right of each
surviving pointer is deleted. This means that every pointer can
act on its own ``reserved'' $n$ sites, i.e. there are no
overlapping reserved areas. Having  the pointer and the reserved
$n$ 2-atom sites, we can effectively address single sites of this
reserved space. This enables us to reduce the number of atoms in
each site in this reserved areas to one and afterwards to empty
the remaining sites that are not reserved by any pointer. In terms
of the operations described above, the protocols is given by a
sequence of the following operators: 1) $U_{1,0}^{0,1}$; 2)
$S_{-1}$, $U_{2,1}^{3,0}$, $E_p$, $U_{2,1}^{3,0}$, and then repeat
this whole step $n-1$ times; 3) $U_{2,1}^{1,2}$; 4) $S_{1}$,
$U_{2,2}^{3,1}$, $E_p$, $U_{3,0}^{1,2}$, and then repeat this
whole step $n-1$ times; 5) $S_1$, $U_{2,0}^{0,2}$,$E_p$,
$U_{2,0}^{1,1}$.

The randomly distributed quantum computers consist of $n$ sites,
all of them with a single atom in state $|a\rangle$, and the
pointer atom in state $|p\rangle$ in the rightmost site (see Fig.\
1). The first atoms store a qubit each, with states
$|{\downarrow}\rangle=|1,0,0\rangle$ and
$|\uparrow\rangle=|0,1,0\rangle$, whereas the pointer atom in
state $|0,0,1\rangle$ carry out the quantum gates.

\begin{figure}[t]
\centering
\includegraphics[width=85mm]{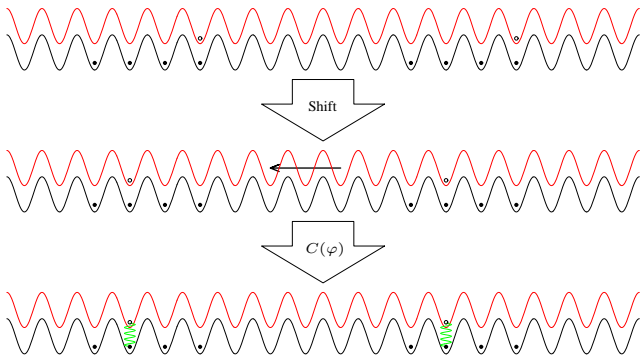}
\label{fig1} \caption{
The levels $a, b$ will store the qubits, and the p--level will
contain the pointer, which can be moved around allowing us to
address single sites, e.g., applying a phase--gate $\varphi$ on an
arbitrary qubit }
\end{figure}

Now we show how to carry out a universal set of quantum gates
using the operations defined above. The idea is to move the
pointer atom to the sites which participate in the quantum gate
and then apply the appropriate operations. The set is composed of
\cite {NiCh}: (a) control-$\pi$ phase gate on two arbitrary
qubits: We first move the pointer to the first site, and apply the
operator $U_{1,1}^{2,0}$. Then we move it to the other site and we
wait until the collisional shift operation $C(\pi)$ is applied.
Finally, we move the pointer back to the first site and apply
again $U_{1,1}^{2,0}$. It is simple to show that this will only
add a $\pi$ phase if both qubits are in the state
$|\uparrow\rangle$. Note that after the first step the pointer
atom is sometimes transferred into the $a$ level. Moving the
pointers to the second qubit and back acts like the identity
operator in those cases; (b) Phase--gate $\varphi$ on an arbitrary
qubit (see Fig. 2): We bring the pointer to the corresponding site
and wait for the appropriate collisional shift operation
$C(\varphi)$; (c) Hadamard gate on an arbitrary qubit: We first
bring the pointer to the site. Then we apply the following
sequence of operations: $V$, $C(\pi)$,$V^\dagger$, and then
$C(\pi/2)$.

Measurement on an arbitrary qubit in the
$\ket{\downarrow},\ket{\uparrow}$ basis can be performed as
follows.  We promote the corresponding atom to the pointer level
provided it is in state $\ket{a}$ i.e., if the qubit is in the
state $\ket{\downarrow}$ . For the measurement, we count the
numbers of atoms in the pointer level (by analyzing the
fluorescence coming from that level) and drop them afterwards.
Note that this occurs in the same way as in usual ensemble quantum
computation \cite{NiCh}, in which we get the global information
about all quantum computers. To save the original pointer from
being emptied we need an extra resting--site, with one atom in the
ground state (for example, the rightmost qubit can be reserved for
this purpose). In summary, we: 1) move the pointer to the
corresponding site and apply $U_{1,1}^{0,2}$; 2) move pointer to
the resting--site and apply $U_{1,1}^{2,0}$ and $U_{1,2}^{2,1}$;
3) count atoms in pointer level and apply $E_p$; 4) apply
$U_{2,0}^{1,1}$. The measured qubit-site is emptied, iff the qubit
was found in state $\ket{\downarrow}$, while the pointer and the
resting-qubit survived unchanged. We can continue by moving the
pointer back to the target qubit, applying $W$ and then repeating
above protocol. The number of atoms counted in the pointer-level
this time is equal to the number of qubits with measured in state
$\ket{\uparrow}$. Alternatively, we can leave out this step
and relate the number of qubits found in $\ket{\downarrow}$ to the
total number of quantum computers in the lattice. This number can
be estimated either by the statistics of the starting distribution
or by measuring  the number of pointers/atoms at the end of the
computation.

So far we have shown how to build a quantum register of $n$
qubits, for any arbitrary integer $n$, and how to perform quantum
computations. Note that in order to prepare the initial state it
is necessary that there are areas in the lattice which have no
defects, i.e. no empty sites nor 1 atom sites. If the number of
such defects well inside the lattice is larger than the number of
1D lattices, then the probability of ending up with at least one
quantum computer will decrease exponentially with $n$, and thus
the method proposed here will not be scalable. In detail, if we
assume that every site of the  lattice is filled independently
with zero, one or two atoms, according to the probabilities
distribution $p_0, p_1 ,p_2 $, then the expected number of quantum
computer of length $n$ in a 1D lattice can be estimated by $N p_1
p_2^{n}=N p_1(1-p_0-p_1)^n$, where the length $N$ of the lattice
has to be much larger than $n$. This quantity decreases
exponentially with $n$ which makes the proposed method not
scalable.

In the following we show how to boost the probability of creating
a quantum computer in the lattice by correcting the defects, i.e.,
making $p_0$ and $p_1$ arbitrarily small. Having this possibility,
we change the probabilities to  $p_0=0$ and $p_1=1/n$. The
resulting expected number of quantum computers  in a lattice of
size $N$ is then given by $N/n (1-1/n)^{n}$, which goes to
$N/(n\cdot e)\sim 1/n$ for large $n$, i.e. our method becomes
scalable. The procedure of correcting the defects will also be
useful if one would like to perform quantum simulations with large
spin chains.

The main idea of the protocol is to first fill all sites which are
empty with one atom coming from a different site, which is
overpopulated. Then, the sites with one atom are filled with
another atom also coming from overpopulated sites (see Fig.\ 3).
Thus, we have to assume that there are as many overpopulated sites
as defects, an achievable requirement for sufficiently high
densities.

\begin{figure}[t]
\centering
\includegraphics[width=85mm]{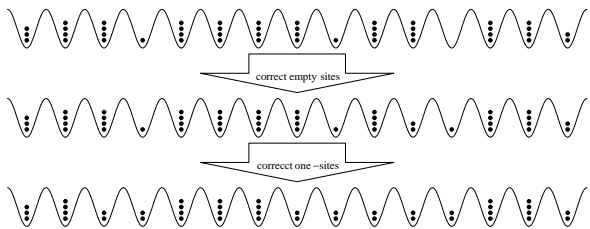}
\label{fig4} \caption{Overpopulated sites are used to first fill
empty sites and then to put two atoms in sites with one atom. }
\end{figure}

First, we reduce all occupation number $>4$ to four \cite{xxx}.
Then, the  protocol starts out by promoting two atoms to the
state $p$ whenever a site has four atoms. Then we shift the
lattice corresponding to level $p$ by a random amount $x$ and try
to deposit one of such atoms in an empty site. The remaining atom
in the $p$ level is thrown away. Note, that for every corrected
defect we lose one atom in this protocol. Losing atoms while
correcting defects is unavoidable, since it is the only way to
reduce the entropy of the state in our setup. We proceed in the
same vein until we make the probability of having sites with no
atoms vanishingly small. In more detail, we apply the following
sequence of operations several times: $U_{4,0}^{2,2}$, $S_x$,
$U_{0,2}^{1,1}$, $S_{-x}$, $U_{4,0}^{2,2}$, $E_p$. With this we
will have filled the empty sites. Now, we can do the same but
replacing $U_{4,0}^{2,2}$ and $U_{0,2}^{1,1}$ by $U_{x,0}^{x-2,2}$
and $U_{1,2}^{2,1}$, so that sites with a single atom get double
occupation. For a finite lattice of length $N$ there are only $N$
different possibilities for the $x$, so the protocol requires at
most $N$ repetitions.

We still need some defects to act as pointers. So we either do not
fill up all the one-atom defects or we have to create new defects.
The latter can be done by first reducing all occupation numbers to
two and then applying a unitary operation that changes
$\ket{2,0,0}$ to the superposition state $\sqrt{\varepsilon}
\ket{1,1,0}+\sqrt{1-\varepsilon} \ket{2,0,0}$ followed by  $E_b$.
With probability $\varepsilon$ a one-atom-site defect is created
out of a 2--atom site.

We have shown that it is possible to perform quantum computations
in optical lattices in the presence of lattice defects and without
the necessity to address single lattice sites, nor to specify the
total number of atoms in the lattice. In practice, a very high
degree of control is required, something which is being achieved
in current experiments with optical lattices. The ideas presented
here not only apply to the field of quantum computation but they
can also be used to prepare and manipulate the states in the
lattice, and to build some prescribed atomic patterns
\cite{CirZo}. Furthermore, all these methods can be generalized in
a straight forward way to 2--dimensional or 3--dimensional
lattices. Finally, note that some of the protocols given here
require a large number of steps, something which is experimentally
demanding. We are currently using the ideas of quantum compression
in order to develop new efficient methods for loading the lattices
\cite{Vol}.

Work supported by EU IST projects, the DFG, and the
Kompetenz\-netz\-werk Quanten\-informations\-verarbeitung der
Bayerischen Staatsregierung.


\begin{thebibliography}{99}
\bibitem{xxx} Alternatively, we can repeat the whole protocol with  occupation
number $4, 5 ,6, \dots$ as overpopulated resource.
\end{thebibliography}
\end{document}